\documentclass[journal]{IEEEtran}

\usepackage{ifpdf}
 \ifpdf
 \else
 \fi

\usepackage{cite}

\ifCLASSINFOpdf
   \usepackage[pdftex]{graphicx}
   \DeclareGraphicsExtensions{.pdf,.jpeg,.png}
\else
   \usepackage[dvips]{graphicx}
   \DeclareGraphicsExtensions{.eps}
\fi

\begin{document}
\title{Discovery of higher order reentrant modes by constructing a cylindrical symmetric ring and post cavity resonator}

\author{Yaohui~Fan,~Zhengyu~Zhang,~Natalia~C.~Carvalho,~Jean-Michel~Le~Floch,~Qingxiao~Shan~and~Michael~E.~Tobar,~\IEEEmembership{Fellow,~IEEE}% <-this % stops a space
\thanks{Manuscript received December 16, 2013.}
\thanks{Y. Fan, N. C. Carvalho, J-M Le Floch and M. E. Tobar are with the ARC Centre of Excellence Engineered Quantum Systems (EQuS), The University of Western Australia, 35 Stirling Highway, CRAWLEY, WA 6009, (corresponding author phone: 61-8-6488-3612; fax: 61-8-6488-1235; e-mail: yaohui.fan@uwa.edu.au).}% <-this % stops a space
\thanks{Z. Zhang is with the Department of Physics, The University of Science and Technology of China (USTC), Hefei 230026, China.}% <-this % stops a space
\thanks{Q. Shan is with College of Mechatronics and Automation, National University of Defense Technology, Changsha 410073, China.}% <-this % stops a space
}

\maketitle

\begin{abstract}
Analysis of the properties of resonant modes in a reentrant cavity structure comprising of a post and a ring was undertaken and verified experimentally. In particular we show the existence of higher order reentrant cavity modes in such a structure. Results show that the new cavity has two re-entrant modes, one which has a better displacement sensitivity than the single post resonator and the other with a reduced sensitivity. The more sensitive mode is better than the single post resonator by a factor of 2 to 1.5 when the gap spacing is below 100 $~\mu$m. This type of cavity has the potential to operate as a highly sensitive transducer for a variety of precision measurement applications, in particular applications which require coupling to more than one sensitive transducer mode.

\end{abstract}


% \fi
\IEEEpeerreviewmaketitle

\section{Introduction}

\IEEEPARstart{R}{ecently}, there has been increased interest in optomechanical and electromechanical systems\cite{Aspelmeyer2013}, which parametrically couple a driven electromagnetic mode to a high-Q, low frequency mechanical mode. Such devices are capable of being used for displacement measurements, sideband cooling, amplification of mechanical motion, and for investigating quantum behavior of macroscopic mechanical resonators\cite{Teufel2011,Teufel2009,Massel2011,Groblacher2009}. A microwave reentrant cavity transducer is such a device, which can provide a very sensitve high-Q microwave mode. The reentrant cavity has been developed in the past for high precision transductance of gravitational wave detectors (highly sensitive and highly massive optomechanical systems)\cite{Linthorne1992,Tobar1993a,Mittoni1993,Blair1995,Pimentel2008a,Aguiar2012}, attempting to measure the standard quantum limit (SQL)\cite{Tobar2000} and investigating the dynamic Casimir effect\cite{Braggio2005a, Giunchi2011}. The key component of the reentrant transducer is a narrow-gap superconducting reentrant cavity, which could achieve high displacement sensitivity of about few hundreds MHz/$\mu$m\cite{Linthorne1992}, and high electrical Q-factor of up to $10^8$ at 1.5 K\cite{Bassan2008}. All the previous work on reentrant cavity resonators discuss single mode devices, in this work we show for the first time that higher order modes exist, which is of general interest as many of these applications would benefit directly by having more than one highly sensitive transducer mode in the cavity.

Reentrant cylindrical cavities have been studied for more than 50 years\cite{Hansen1939,Hahn1941,Fujisawa1958,Jaworski1978} and have been implemented in multiple ways\cite{Humphries1986,Kaczkowski1980a,Kaczkowski1980b,Baker-Jarvis1998,Barroso2004}. Previously, we constructed a tunable reentrant cavity, which consists of an empty cylindrical cavity and a cylindrical post in the centre of the cavity\cite{LeFloch2013}. The position of the cylindrical post was adjustable along the cylindrical axis of the resonator by a fine screw mechanism, which offers a rotation-free translation of the post. By gradually screwing the post into the cavity the resonant mode was transformed from the standard $TM_{0,1,0}$ mode in an empty cylinder, to the fundamental TM mode of reentrant cavity. Since the field structure of the mode was dramatically transformed, a very large tuning range was achieved from 22 GHz down to about 2 GHz. Also, as the post was inserted and became close to the adjacent cavity wall, the electric field was confined within the capacitive gap region formed by the post and the wall leading to a very high displacement sensitivity of around $\sim$350 MHz/$\mu$m at 10 $\mu$m gap size. 

\begin{figure}[!t]
	\centering
		\includegraphics[width=0.48\textwidth]{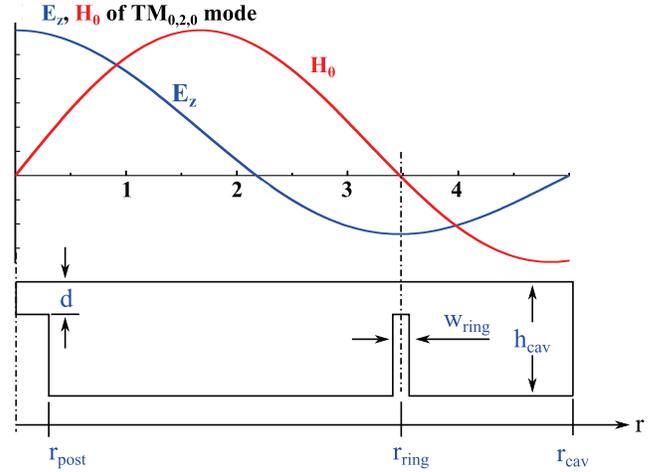}
	\caption[Fig.1]{Above: The relative field amplitude of the $TM_{0,2,0}$ mode in a hollow cylindrical cavity. Below: The cylindrical post and an additional ring of the Reentrant-Ring cavity is positioned at the two anti-nodes of $E_{z}$ field as shown to create the reentrant cavity mode. In this work the height of the ring is kept same as the central post.}
	\label{fig:fig1}
\end{figure}

In a similar way it should be possible to create higher order reentrant cavity modes based on the $TM_{0,n,0}$ mode of a cylindrical cavity ($n>1$). In this paper we show the existence of the next order mode with $n=2$. The azimuthal magnetic field, $H_{\theta}$, of $TM_{0,2,0}$ mode includes two nodes along the radial axis of the cavity, while the $E_{z}$ field has two anti-nodes (see the top curve of Fig.~\ref{fig:fig1}). To create the higher order mode we conceive the concept of the Reentrant-Ring (RR) cavity, while the common reentrant cylindrical cavity we refer to as a Single-Post (SP) cavity. The geometry of the RR cavity is shown schematically in the bottom plot of Fig.~\ref{fig:fig1}, which consists of a central post and an additional ring positioned at the two anti-nodes of $E_{z}$, $r=0$ and $r=r_{ring}$, respectively. In this work the heights of the central post and the ring are kept the same for simplicity. The cavity was simulated using the finite-element method (FEM) and in comparison with the SP reentrant cavity (of the same size) the RR cavity is shown to have a higher displacement sensitivity. Several cavity configurations were examined experimentally, with consistent results verified from the finite-element simulations.

\section{Finite element analysis of reentrant cavities}

\begin{figure}[!t]
	\centering
		\includegraphics[width=0.48\textwidth]{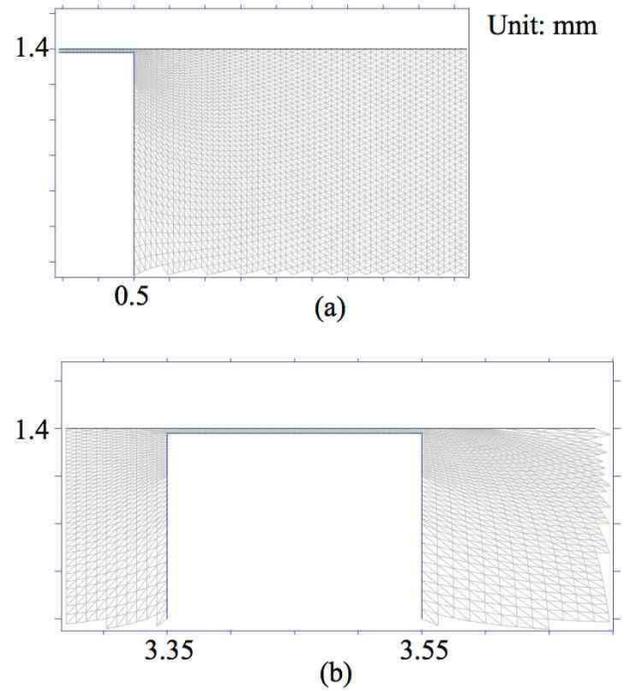}
	\caption[Fig.2]{Mesh plots of a RR cavity model with the aspect ratio of the cavity height to the gap size equals to 280 (5 $\mu$m gap). (a) gap region at the top of the central post; (b) gap region at the top of the ring.}
	\label{fig:fig2}
\end{figure}

Since the RR cavity is an axisymmetric structure it was analyzed using a 2D finite element model. Poisson Superfish software~\cite{Billen1997} was used for the modeling. The meshing of such a cavity is non-trivial, as the cavity has a gap of only a few micrometers with an extreme aspect ratio of the cavity's overall size to the gap spacing, which can be greater than 100. We cannot create a too small mesh size for the whole cavity because of the memory required and CPU time to calculate and store the solutions. Therefore, it is necessary to mesh differently depending on this aspect ratio. In the meantime, meshes were optimized according to the ratio of the electric and magnetic stored-energy integrals of the entire cavity. The difference between the two stored-energy integrals depends on the mesh size. Finer meshes result in better agreement. Fig. \ref{fig:fig2} shows an example of such mesh plots around the two gap regions in a RR cavity.

\begin{figure}[!t]
	\centering
		\includegraphics[width=0.48\textwidth]{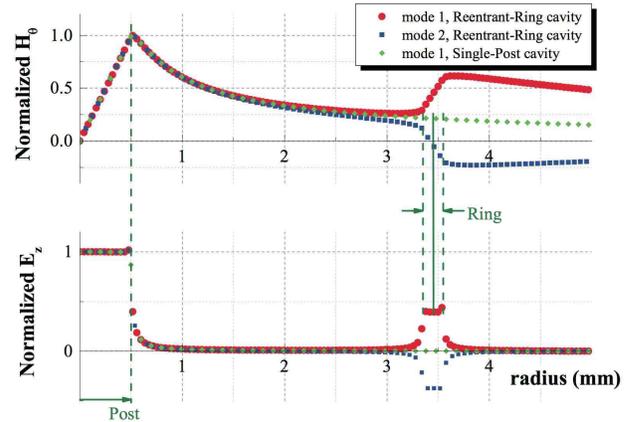}
	\caption[Fig.3]{Comparison of $H_\theta$ and $E_z$ field distributions within gap region between the SP cavity mode and the RR cavity modes. Both cavites have the same gap size of 20 $\mu$m. \textit{Mode 1} and \textit{mode 2} are the fundamental TM mode ($n = 1$) and the first high-order TM mode ($n = 2$), respectively.}
	\label{fig:fig3}
\end{figure}

We found the first two TM modes of the RR cavity had analogous field patterns to the fundamental TM mode of the SP cavity. All three modes have the same dominant field components: the axial electric field, $E_z$, and the azimuthal magnetic field, $H_\theta$. Fig. \ref{fig:fig3} shows comparisons of these field distributions within the gap regions (which in this case is 20 $\mu$m).  The terms \textit{mode 1} and \textit{mode 2} in the figure represent the fundamental TM mode ($n = 1$) and the first high-order TM mode ($n = 2$) , respectively. One can see that most of electric energy is retained in the gap regions, with the stronger $E_z$ field in the central post gap region. However, the electric filling factors between this two gap regions are determined by the cavity dimensions. From the field plots we can see by inserting the ring, two TM modes are created. Within the ring both modes have nearly identical $H_{\theta}$ and $E_z$ field patterns, with intense $H_{\theta}$ field around the central post region. However, in the exterior ring region, the $H_{\theta}$ and $E_z$ fields of the two modes are orientated in opposite directions, with the $n = 2$ mode changing sign. It is noted that the field components of the SP cavity are approximately equal to the average of the field of the two modes in the RR cavity.

\begin{figure}[!t]
	\centering
		\includegraphics[width=0.48\textwidth]{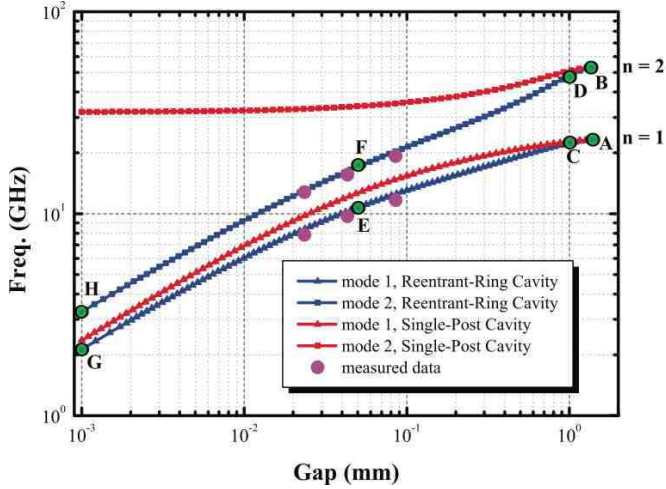}
	\caption[Fig.4]{Resonant frequency as a function of gap size calculated using finite-element method. The blue curves are the fundamental mode and the first high-order mode of the Reentrant-Ring cavity, and the red curves are resonant modes of the Single-Post cavity. Points A-H indicate evolution of the \textit{mode 1} and \textit{mode 2} of the RR cavity, with density plots shown in Fig.\ref{fig:fig5}. Experimental point are displayed for three cavities manufactured with three different gap spacings, which verify the validity of the simulations. }
	\label{fig:fig4}
\end{figure}

\begin{figure*}[!t]
	\centering
		\includegraphics[width=0.96\textwidth]{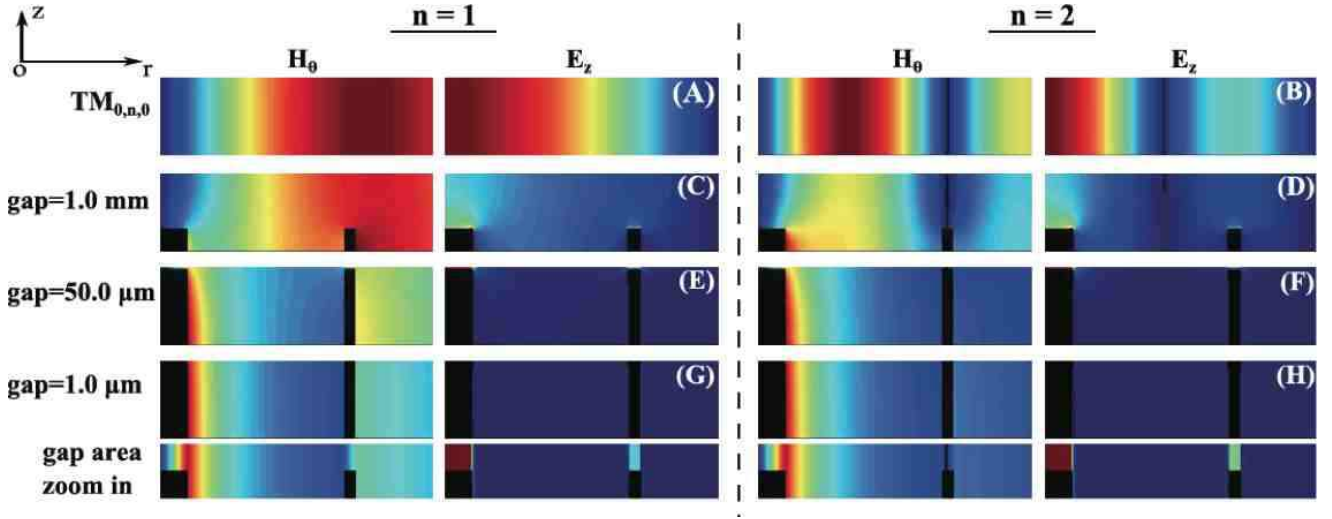}
	\caption[Fig.5]{Density plots of $H_\theta$ and $E_z$ fields in half of the cylindrical cavity, for the modes shown in Fig.\ref{fig:fig4} as a function of gap size. On the left side, \textit{'n=1'} indicates the $TM_{0,1,0}$ mode and the fundamental TM mode (\textit{mode 1}). \textit{'n=2'} on the right side represents the $TM_{0,2,0}$ mode and the first high-order mode (\textit{mode 2}).}
	\label{fig:fig5}
\end{figure*}

We have calculated the resonant mode properties with respect to various gap sizes from one micrometer to above one millimeter. Fig.~\ref{fig:fig4} shows resonant frequency tuning curves comparison between the RR cavity modes and the SP cavity modes. At the rightmost of Fig. \ref{fig:fig4}, the modes converge and correspond to the $TM_{0,1,0}$ mode and $TM_{0,2,0}$ mode of the empty cylinder, respectively. All modes have a significant frequency tuning with displacement, except for \textit{mode 2} of the SP cavity, which transforms to a standard co-axial mode as the post and ring is pushed into the cavity. Fig.\ref{fig:fig5} illustrates the evolution of the field patterns as the gap size changes from over 1 mm down to 1 $\mu$m. It is clear that three of the modes are transformed from the $TM_{0,n,0}$ modes of the empty cylinder to the TM modes of reentrant cavity during the process.

\begin{figure}[!t]
	\centering
		\includegraphics[width=0.48\textwidth]{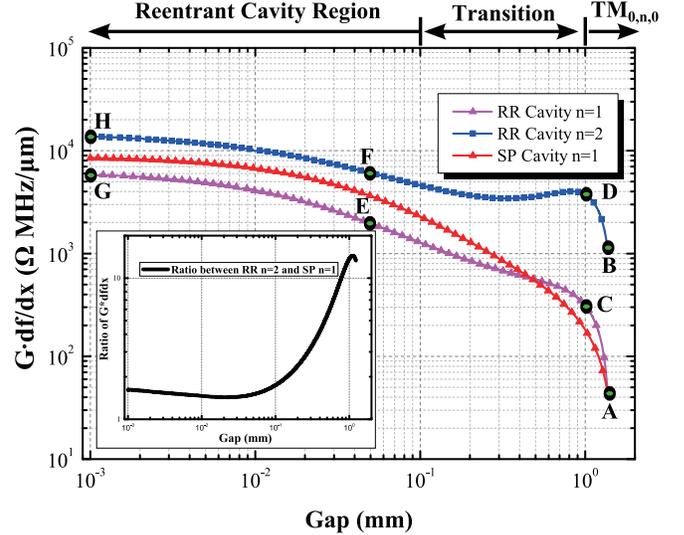}
	\caption[Fig.6]{$G \cdot \frac{df}{dx}$ as a function of gap size calculated using the finite-element method. Points A-H indicate the evolution of \textit{mode 1} and \textit{mode 2} of the RR cavity, with density plots shown in Fig.\ref{fig:fig5}. The modes behave like re-entrant modes below 100 $\mu$m gap size and perturbed $TM_{0,n,0}$ modes above 1 mm, while in between is the transition region. Inset: Ratio of $G \cdot \frac{df}{dx}$ between the RR cavity ($n$ = 2) and the SP cavity ($n$ = 1), showing the factor of improvement in sensitivity, which is 1.6 at 1 $\mu$m, 1.5 at 10 $\mu$m and 2.0 at 100 $\mu$m. }
	\label{fig:fig6}
\end{figure}

One important application of the reentrant cavity is that of a highly sensitive transducer to measure very weak forces and tiny displacements. To achieve maximum displacement sensitivity the product $Q \cdot \frac{df}{dx}$ must be maximised, where $Q$ is the Q-factor of the resonant cavity and $\frac{df}{dx}$ is frequency tuning coefficient in Hz per metre~\cite{Tobar1993a}. The Q-factor is determined by the geometric factor $G$ and the surface resistance $R_s$ of the material through the relation $Q=G/R_s$. Thus, to optimize the transducer independent of material properties we optimise the product of $G \cdot \frac{df}{dx}$.  Fig.~\ref{fig:fig6} shows the calculated $G \cdot \frac{df}{dx}$ as a function of gap size. It is evident that \textit{mode 2} of the Reentrant-Ring cavity has the largest $G \cdot \frac{df}{dx}$ value, which means this RR cavity mode has the potential to act as a better transducer in comparison with the common SP reentrant cavity structure. It is also evident that \textit{mode 1} has worse sensitivity, however the average is close to the sensitivity of the SP resonator in a similar way that the $H_{\theta}$ and $E_z$ fields of the two modes are approximately the average of the SP cavity.

\begin{figure}[!t]
	\centering
		\includegraphics[width=0.48\textwidth]{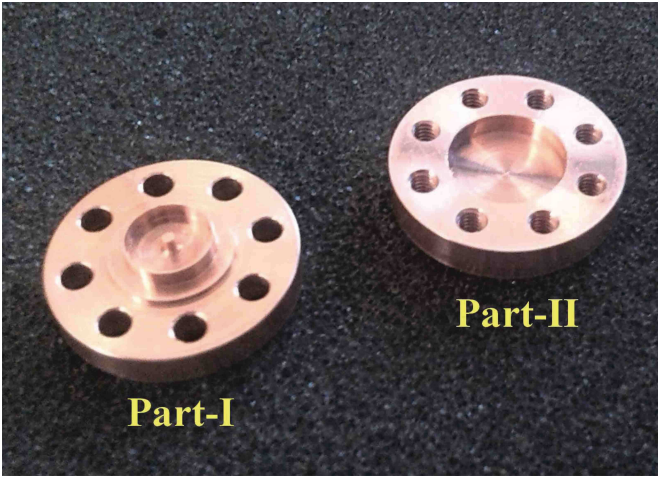}
	\caption[Fig.7]{Photograph of a copper Reentrant-Ring cavity. Part-I consists of the central post and the ring. Part-II is an empty cylindrical cavity. The cavity is 10.0 mm in diameter and 1.5 mm in height. The ring is 6.7 mm in diameter, and 0.2 mm in thickness. The diameter of the central post is 1.0 mm. Part-I and Part-II are assembled by screws.}
	\label{fig:fig7}
\end{figure}

%\IEEEpubidadjcol

\section{Experimental results}
\begin{figure}[!t]
	\centering
		\includegraphics[width=0.48\textwidth]{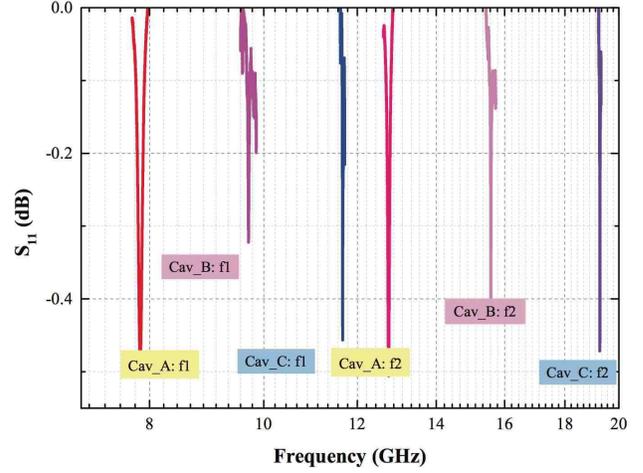}
	\caption[Fig.8]{The measured absolute values of $S_{11}$  of \textsl{mode1} ($f_1$) and \textsl{mode2} ($f_1$) for the three sets of cavities of different gap spacings, with the corresponding resonant frequencies tabulated in Table~\ref{tab:Table1}.}
	\label{fig:fig8}
\end{figure}

\begin{figure}[!t]
	\centering
		\includegraphics[width=0.48\textwidth]{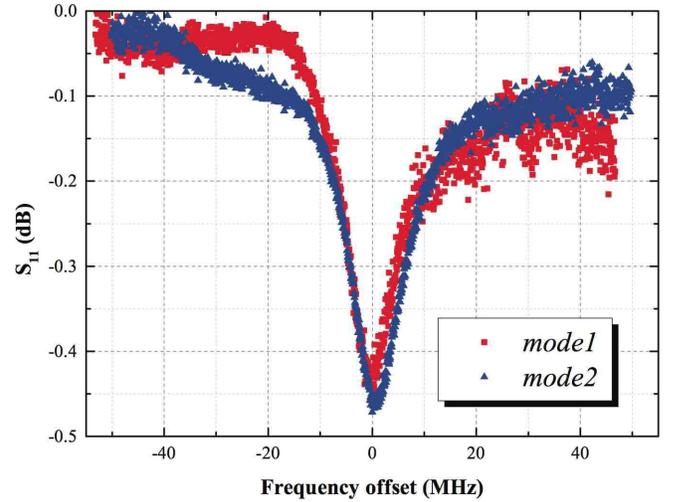}
	\caption[Fig.9]{The measured absolute values of $S_{11}$  of \textsl{mode1} $f_1$=11.662 GHz and \textsl{mode2} $f_2$=19.27 GHz for the 80 $\mu$m gap RR cavity. The corresponding Q-factors are listed in Table~\ref{tab:Table2}. }
	\label{fig:fig9}
\end{figure}

\begin{table*}[!t]
\renewcommand{\arraystretch}{1.3}
	\centering
	\caption{\label{tab:Table1}Predicted frequencies of \textsl{mode 1} and \textsl{mode 2} in comparison with experimental data. Units: Gap in $\mu m$, and Frequency in GHz.}
	\begin{tabular} {c c c c c c c c}
			\hline\hline
			  & $Gap_{sim}$ & $f_{1sim}$ & $f_{1meas}$ & error & $f_{2sim}$ & $f_{2meas}$ & error \\ \hline
			Cav\_A & 23.5 & 8.073 & 7.855 & 2.28$\%$ & 12.485 & 12.758 & 2.18$\%$ \\ 
			Cav\_B & 43.2 & 9.803 & 9.706 & 0.78$\%$ & 15.498 & 15.578 & 0.75$\%$ \\
			Cav\_C & 85.6 & 11.767 & 11.662 & 0.89$\%$ & 19.088 & 19.270 &  0.96$\%$ \\
			\hline\hline
		\end{tabular}
\end{table*}
The properties of the resonant modes for both the RR and SP cavities were measured using a vector network analyzer to acquire the complex values of $S_{11}$ in reflection~\cite{Kajfez1984}, with a loop probe inserted through a 1.0 mm diameter hole from the side wall of the cavity to coupling to the $H_{\theta}$ field component. The absolute values of magnitude of $S_{11}$ for mode 1 and 2 for the three RR cavities are shown in  Fig.\ref{fig:fig8}, with a close up of the modes in the 80 $\mu$m gap RR cavity shown in  Fig.\ref{fig:fig9}. To calculate the unloaded Q-factor, a Q-circle plot was fitted to the data to determine the resonance frequency, bandwidth and coupling. The measured frequencies of \textsl{mode 1} ($f_1$) and \textsl{mode 2} ($f_2$) are compared to the predicted values, which are tabulated in Table~\ref{tab:Table1} and shown pictorially in Fig.~\ref{fig:fig4}. The observed frequencies are very close to the predicted frequencies with maximum deviation about 2 percent.

From the Q-factor measurements the material surface resistances, $R_s$, and conductivity, $\sigma$, of the copper may be calculated from the following relation:

\begin{equation}
	R_s=\frac{G}{Q} = \sqrt{\frac{\pi \mu_0 f}{\sigma}}.
	\label{eqn:Rs}
\end{equation}

Here $\mu_0$ is the permeability of free space and $f$ is the resonance frequency of the mode under consideration.
For example, Table~\ref{tab:Table2} shows the calculated material properties from the properties of the modes measured in Cavity C, which has a gap of 85.6 $\mu m$, using the relations given in equation~(\ref{eqn:Rs}). The calculated values of conductivity are shown in the fourth column of table~\ref{tab:Table2} and are consistent with typical known values of normal copper. They also reveal a constant value of conductivity of $3.7\times10^7$ siemens per meter at 11.7 and 19.3 GHz, revealing frequency independence, which is also well knowm for normal metals like copper.

\begin{table}[!t]
\renewcommand{\arraystretch}{1.3}
	\centering
	\caption{\label{tab:Table2}Measured surface resistance and conductivity from the resonant modes in the $80\ \mu m$ RR cavity.}
		\begin{tabular} {c c c c c}
			\hline\hline
				 & $Q$ & $G (\Omega)$ & $R_s (m \Omega)$ & $\sigma_{copper} (S/m)$ \\ \hline
				$mode 1$ & 1022.8 & 35.95 & 35.15 & 3.72e7 \\ 
				$mode 2$ & 1406.5 & 64.23 & 45.67 & 3.65e7 \\
			\hline\hline	
		\end{tabular} 
\end{table}

\section{Conclusion}
In summary, we have presented the new concept of a Reentrant-Ring cavity and verified the modes through numerical finite-element modelling and comparison with experiments. We found the existence of higher order reentrant cavity modes and in principle a multi-ring structure could be engineered to create a cavity with multiple sensitive modes, with the frequencies below that of the normal TE and TM modes of a cylindrical cavity. For our $n=2$ structure results show that such a cavity has two modes, one with better sensitivity in comparison with the common single post reentrant cylindrical cavity, while the other was less sensitive. This type of cavity has room for further optimization and should be very promising for a variety of precision measurement applications. For example, most hybrid quantum engineered schemes now utilise more than one cavity mode, and need high tunability. Moreover, it is important that the sensitive modes are far detuned from all other spurious modes within the system, which is another promising feature of this type of cavity.

\section*{Acknowledgment}

This work has been supported by the Australian Research Council Grants FL0992016 and CE110001013.

\ifCLASSOPTIONcaptionsoff
  \newpage
\fi



%\bibliographystyle{IEEEtran}

%%%\bibliographystyle{IEEEtran}
%%%\bibliography{IEEEabrv,paper}
%%%

\end{document}